\documentclass[12pt,preprint]{aastex}
\newcommand{\etal}{{\it et al.}}
\newcommand{\beq}{\begin{equation}}
\newcommand{\eeq}{\end{equation}}
\newcommand{\ben}{\begin{eqnarray}}
\newcommand{\een}{\end{eqnarray}}
\begin{document}
\title{Formalism for the Subhalo Mass Function in the Tidal-limit
Approximation}
\author{\sc Jounghun Lee}
\affil{Department of Physics, The University of Tokyo, Tokyo 113-0033, Japan}
\email{lee@utap.phys.s.u-tokyo.ac.jp}
\received{2003 October 30}
\accepted{2003 ???}
\begin{abstract} 
We present a theoretical formalism by which the global and the local
mass functions of dark matter substructures (dark subhalos) 
can be analytically estimated. The {\it global} subhalo mass 
function is defined to give the total number density of dark
subhalos in the universe as a function of mass, while the 
{\it local} subhalo mass function counts only those subhalos 
included in one individual host halo.  We develop our formalism by
modifying the Press-Schechter theory to incorporate the followings:  
(i) the internal structure of dark halos; (ii) the correlations between 
the halos and the subhalos; (iii) the subhalo mass-loss effect driven by 
the tidal forces.   We find that the resulting (cumulative) subhalo mass 
function is close to a power law with the slope of $\sim -1$, that the 
subhalos contribute approximately $10 \%$ of the total mass,  and that 
the tidal stripping effect changes the subhalo mass function
self-similarly, all consistent with recent numerical detections.  
 
\end{abstract} 
\section{INTRODUCTION}

The dark halo substructures ({\it dark subhalos}) are the dynamically 
distinct, self-bound objects in virialized dark matter halos.  
The presence of substructures in the dark matter halos is a generic 
picture of the cold dark matter (CDM) cosmology.  Recent numerical 
simulations of ultra-high resolution indeed confirmed that the dark 
halos are not smooth structureless objects but clumpy systems marked 
by a wealth of substructures 
\citep{tor-etal98,kly-etal99,oka-hab99,ghi-etal00,       
spr-etal02,zha-etal02,delucia-etal03,hay-etal03,zen-bul03}.  

Recently, the mass function of dark subhalos has drawn sharp attentions 
\citep{fuj-etal02,she03,bla03} especially because of its connection to 
the galaxy luminosity function. 
Yet, it is not an easy task to derive the subhalo mass function either 
in numerical or analytical ways.  The numerical approach to the subhalo
mass function using N-body simulations still suffers from resolution
effects related to the so called over-merging problem
\citep{kly-etal99}.   Even recently  available ultra-high resolution 
simulations are capable of producing only the local subhalo mass
function, i.e., the mass function of the subhalos within one individual 
dark halo \citep{oka-hab99,ghi-etal00}.  Given the importance of the 
subhalo mass function as a clue to understanding of the galaxy
luminosity function, however, what is also desired is the
global subhalo mass function, i.e., the mass function of all the
subhalos in the universe, irrespective of the host halos. 

As for the analytic approach, the hindrance is the complexity of 
the subhalo evolution.  
For the mass function of dark halos,  we already have a remarkably 
successful theory developed by \citet[][hereafter PS]{pre-sch74}.   
The principle of the PS theory is this:  the formation and evolution of 
dark halos can be traced by the linear theory, assuming (i) dark halos 
have no internal structure;  (ii) dark halos form independently of their 
surroundings; (iii) dark halos do not lose mass in the evolution but 
only hierarchically merge via gravity. 
Unlike the case of the halo mass function, however, the subhalo 
mass function cannot be derived under such simple assumptions. The subhalos
are, by definitions, the internal structures of the halos, being placed 
in highly dense surroundings, and thus the formation and evolution of 
the subhalos must depend strongly on their surroundings.  
Among the various consequences from the surrounding 
influences, the most significant one is the subhalo mass-loss: 
the subhalos do not only gravitationally merge but also get disrupted 
or at least lose considerable amount of their mass through the 
interaction with the surroundings.  In fact, it has been demonstrated 
by several N-body simulations that the subhalos lose most of their mass 
throughout the evolution, contributing after all only $10-15\%$ of the 
total mass of the host halos \citep[e.g.,][]{tor-etal98}.  
There are three different processes that can drive the subhalos to lose
mass: the global tides generated by the host halos, the dynamical frictions, 
and the close-encounters with the other subhalos.  Apparently, the subhalo 
mass-loss is quite a complicated process, so that it would be
practically impossible to take into account its effect fully in  
deriving the subhalo mass function analytically.  That was why all
previous analytic approaches had to make the unrealistic assumption that
the subhalos do not lose mass during the evolution 
\citep{fuj-etal02,she03,bla03}. 

However, the mass-loss phenomenon is the most essential feature of the
subhalo evolution, which must be taken into account in order to
estimate the subhalo mass function in any realistic sense.  
Here we attempt for the first time to estimate both the global and the
local subhalo mass functions with the subhalo mass-loss effect taken 
into account.  To make the theory analytically tractable, we still make 
some simplified assumptions that the subhalo mass-loss is mainly driven 
by the global tides, and that the condition for a subhalo to survive 
the global tides is a simple function of the distance from its host
halo.  

\section{FORMALISM}

The (differential) global subhalo mass function, 
$\frac{dN(M_{s})}{d\ln M_{s}}d\ln M_{s}$, is defined as the number
density of subhalos in logarithmic mass range  
$[\ln M_{s},\ln M_{s}+d\ln M_{s}]$. To estimate it, we assume the following. 
\begin{itemize}
\item[(i)] 
     The gravitational collapse process to form dark matter halos 
     follows the Top-hat spherical dynamics, according to which a dark 
     halo of mass $M$ forms at redshift $z$ if the density contrast  
     $\delta$ ($\delta \equiv \Delta\rho/\bar{\rho}$, $\bar{\rho}$: the mean 
     mass density of the universe) of a Lagrangian region in the linear
     density field smoothed on mass scale of $M$ satisfies the gravitational
     collapse condition of $\delta = \delta_{c}(z)$ where $\delta_{c}(z)$ 
     is the critical density contrast at redshift $z$ whose value
     depends on cosmology \citep{kit-sut96}. 
     For a flat universe of closure density, 
     $\delta_{c}(z) \approx 1.68(1+z)$ \citep{gun-got72}. 
\item[(ii)]
     The mass function of dark halos is well evaluated by the PS theory as  
\begin{equation}
\label{eqn:psmf}
\frac{dN(M)}{d\ln M} = \sqrt{\frac{2}{\pi}}\frac{\bar{\rho}}{M}
\left|\frac{d\ln\sigma}{d\ln M}\right|\nu
\exp\left(-\frac{\nu^{2}}{2}\right), 
\end{equation}
where $\nu \equiv \delta_{c}/\sigma(M)$ and $\sigma(M)$ is the rms 
density fluctuation of the linear density field on mass scale of $M$.  
\item[(iii)] 
     A dark halo hosts a multiple of subhalos, each of which  
     rotates on a circular orbit. The global tidal 
     field of the host halo strips the subhalos, which drives the
     subhalos to either get completely destroyed or survive but with
     reduced mass.  A subhalo of initial mass $M_{1}$ before 
     the tidal stripping effect at an initial distance $r$ from its host 
     halo of mass $M_{2}$ eventually survives the tidal stripping effect, 
     ending up with having reduced mass of $M_{s}$($< M_{1}$), 
     if $M_{1}$, $M_{2}$, and $r$, satisfy the condition 
\begin{equation}
\label{eqn:tsc}
M_{s} = c_{t}\left(\frac{4\pi}{3}\bar{\rho}\right)
\left(\frac{M_{1}}{M_{2}}\right)r^{3},
\end{equation}
     where the proportionality constant $c_{t}$ is a free parameter. 
\item [(iv)]
     The spatial distribution of the subhalos inside their host halo  
     follows that of the dark matter particles, i.e., the profile  
     given by Navarro, Frenk, \& White (1996) (hereafter, NFW):
\begin{equation}
\label{eqn:nfw}
P_{M_{2}}(r) \propto \frac{1}{(r/r_{s})(1 + r/r_{s})^{2}}
\end{equation} 
where  $r_{s}$ is the scale radius.      
\end{itemize}

It is worth noting that equation (\ref{eqn:tsc}) is reminiscent of the 
familiar tidal-limit approximation, according to which a subhalo in the 
tidal field loses all mass beyond its tidal radius, $r_{t}$. 
We set $c_{t}$ as a free parameter since its precise value
depends on underlying assumptions: 
if the halo potentials can be approximated as point mass 
and the linear size of the subhalo is much smaller than the distance 
from the host halo, the tidal radius has a simple expression of 
$r_{t} = (c_{t}M_{1}/M_{2})^{1/3}r$ with $c_{t}=1/2$ 
(the Roche limit); if the effect of the centrifugal force is taken into
account, then $c_{t} = 1/3$ (the Jacobi limit);  
if the halos are treated more realistically as the extended mass
profiles, then the tidal radius has a more general expression 
\citep[see, e.g., ][]{tor-etal98}. 
Although equation (\ref{eqn:tsc}) is an obvious oversimplification 
of real tidal mass-loss process \citep{hay-etal03}, there are numerical 
clues that the fraction of the survival subhalos has a strong correlation 
with the distance from the host halos \citep{oka-hab99}. It implies 
that the tidal survival condition should depend on the subhalo orbital 
distance as well as the host halo and the subhalo mass. 
Therefore, equation (\ref{eqn:tsc}) may be the simplest possible choice 
for the functional form of the tidal survival condition.  
  
Let us first consider $P_{M_{2}}(M_{1};r)$, the conditional probability 
that a subhalo has an initial mass greater than $M_{1}$ (before the tidal
mass-loss) provided that it rotates upon a host halo of mass $M_{2}$ 
at a distance $r$. According to the hypothesis (i), 
$P_{M_{2}}(M_{1};r) = P_{\delta_{2}=\delta_{c}}[\delta_{1}(r)\ge\delta_{c}]$  
where $\delta_{i}$ (with $i=1,2$) is the density contrast of a 
Lagrangian region in the linear density field smoothed on mass scale
$M_{i}$, and $\delta_{ci}$ is the critical value of $\delta_{i}$. 
The conditional probability 
$P_{\delta_{2}=\delta_{c}}[\delta_{1}(r)\ge\delta_{c}]$ can be 
computed from the Gaussian probability density distribution with 
the help of the Bayes theorem.  For the case of the sharp k-space
filter, it has the following simple analytic form \citep{yan-etal96}:
\begin{equation}  
\label{eqn:conpm1}
P_{M_{2}}(M_{1};r) = \frac{1}{\sqrt{2\pi}}\int_{\beta}^{\infty}
\exp\left(-\frac{x^{2}}{2}\right)dx, \quad {\rm with} \quad
\beta \equiv \frac{1}{\sqrt{1-\gamma^{2}}}
\left(\frac{\delta_{c1}}{\sigma_{1}}\right)
\left[1 - \frac{\delta_{c2}}{\delta_{c1}}
\frac{\sigma^{2}_{c}}{\sigma^{2}_{2}}\right].  
\end{equation}
Here $\sigma^{2}_{i}$ (with $i=1,2$) and $\sigma^{2}_{c}(r)$ are  
the mass variance of the linear density field on mass scale of $M_{i}$, 
and the linear density cross correlation, respectively, given as   
\begin{equation} 
\label{eqn:sigmas}
\sigma^{2}_{i} = \int_{-\infty}^{\ln (k_{ci})}\Delta^{2}(k)d\ln k, 
\qquad 
\sigma^{2}_{c} = \int_{-\infty}^{\ln(k_{c2})}\Delta^{2}(k)
\frac{\sin kr}{kr}d\ln k
\end{equation}
where $\Delta_{k}$ is the dimensionless power spectrum of the 
linear density field,  
$\gamma \equiv \sigma^{2}_{c}/(\sigma_{1}\sigma_{2})$, 
and the integral upper limit $k_{ci}$ is related to 
$M_{i}$ by $k_{ci} = (6\pi^{2}\bar{\rho}/M_{i})^{1/3}$. 
In fact, it was \citet{yan-etal96} who first incorporated the spatial 
correlations between the dark halos themselves into the PS theory. 
\citet{fuj-etal02} used the formalism of \citet{yan-etal96} to estimate 
the local subhalo mass function with the spatial correlations between 
the host halos and the subhalos taken into account. 
However, \citet{fuj-etal02} assumed that the spatialdistribution of 
the subhalos is uniform, and averaged $P_{M_{2}}(M_{1};r)$ over $r$
without taking into account the tidal mass-loss and its correlation with $r$.  

Next, let us consider $P_{M_{2}}(r)$, the probability of finding a subhalo 
in a spherical shell of radius $r$ and thickness $dr$ around a host halo
of mass $M_{2}$ in the Lagrangian space.  According to the hypothesis
(iv), it can be written as 
\begin{equation}
\label{eqn:conpr}
P_{M_{2}}(r) = \frac{A}{(r/r_{s})(1 + r/r_{s})^{2}}4\pi r^{2}dr,  \quad 
{\rm with} \quad
A \equiv \frac{1}{4\pi r^{3}_{s}[\ln(1+\frac{R_{2}}{r_{s}})-
\frac{R_{2}}{r_{s} + R_{2}}]}.
\end{equation}
where $R_{2}$ is the virial radius of $M_{2}$, and the amplitude $A$ is 
determined to satisfy the normalization constraint of 
$\int_{0}^{R_{2}}P_{M_{2}}(r)=1$.   We approximate $R_{2}$ by the Top-hat 
radius of $M_{2}$, and adopt the empirical relation for $r_{s}$ 
proposed by \citet{kly-etal99}:  
\begin{equation} 
R_{2} = \left[\frac{3M_{2}}{4\pi\bar{\rho}}\right]^{1/3}, \qquad
r_{s} = \frac{R_{2}}{124}\left[\frac{M_{2}}{h^{-1}M_{\odot}}\right]^{0.084}.
\end{equation}
It is also worth mentioning here that $\delta$,$R_{2}$, $r$, and $r_{s}$
are all measured in the Lagrangian space where the density field is 
still Gaussian. 

The joint conditional distribution, $P_{M_{2}}(r,M_{1})$, i.e., the
probability of finding a subhalo of mass greater than $M_{1}$ at a
distance $r$ from a host halo of mass $M_{2}$, can be derived from equations
(\ref{eqn:conpm1}) and (\ref{eqn:conpr}) by using the Bayes theorem:
$P_{M_{2}}(M_{1},r) = P_{M_{2}}(M_{1};r)P_{M_{2}}(r)$. 
The partial derivative, $\partial P_{M_{2}}(r,M_{1})/\partial M_{1}$, 
is proportional to the fraction of the host halo volume occupied by those
subhalos of mass $M_{1}$ at a distance $r$, $f_{M_2}(M_1,r)$, 
such that 
$f_{M_2}(M_1,r) = (M_{2}/\bar{\rho})
\vert \partial P_{M_{2}}(r,M_{1})/\partial M_{1}\vert$ 
where the proportionality factor $(M_{2}/\bar{\rho})$ is nothing but the 
average volume of the host halo.  Following the familiar 
PS-like approach,  the initial number density of the subhalos of mass 
$M_{1}$ at a distance $r$ inside a host halo of mass $M_{2}$ equals  
$f_{M_2}(M_{1},r)$ divided by the average volume of the subhalo  
$M_{1}/\bar{\rho}$ such that 
$dN_{M_{2}}(r,M_{1})/(4\pi r^{2}dr~dM_{1}) = 
2\left(\bar{\rho}/M_{1}\right)f_{M_2}(M_1)$ where  the factor $2$ is 
the normalization constant introduced by PS 
\citep[see also,][]{pea-hea90,bon-etal91,jed95,yan-etal96,fuj-etal02}. 
We end up with the following expression: 
\begin{equation}
\label{eqn:cond}
\frac{d^{2}N_{M_{2}}(r,M_{1})}{4\pi r^{2}dr~d\ln\tilde{M}_{1}} =  
\sqrt{\frac{2}{\pi}}\left(\frac{1}{\tilde{M}_{1}}\right)
\left|\frac{d\ln\sigma_{1}}{d\ln\tilde{M}_{1}}\right|
\left(\frac{\xi}{1-\gamma^2}\right)\frac{A}{(r/r_{s})(1 + r/r_{s})^{2}}
\exp\left(-\frac{\xi^{2}}{2}\right).
\end{equation} 
where $\tilde{M_{1}} \equiv M_{1}/M_{2}$ and 
$\xi \equiv \frac{\nu_{1}}{\sqrt{1-\gamma^{2}}}
\left(1-\frac{\nu_{1}}{\nu_{2}}\gamma\right)$ with 
$\nu_{i} \equiv \delta_{ci}/\sigma_{i}$ for $i=1,2$.
Equation (\ref{eqn:cond}) is the {\it conditional} mass and spatial 
distribution of the subhalos provided that they are included in 
the host halos of mass $M_{2}$ before the tidal mass-loss.  
If the subhalo mass were conserved, the local subhalo mass function 
would be computed simply by integrating equation (\ref{eqn:cond}) 
over $r$.  However, the subhalos in reality either get destroyed or 
lose mass, so that the subhalo mass function cannot be simply obtained 
in that way. Notwithstanding, the reduced mass of survived subhalos 
can be determined from the informations of $M_{1}$, $M_{2}$, and $r$ 
according to the hypothesis (iii). 

Hence,  we find the local subhalo mass function after the tidal mass
loss as 
\begin{equation}
\label{eqn:localmf}
\frac{dN_{M_{2}}(M_{s})}{d\ln M_{s}}d\ln M_{s} 
= \int_{r_{c}}^{R_{2}}4\pi r^{2}dr\int_{-\infty}^{0}
d\ln \tilde{M}_{1}~\frac{dN_{M_{2}}(r,M_{1})}
{4\pi r^{2}dr d\ln \tilde{M}_{1}}~\delta_D
\left(c_{t}\frac{4\pi}{3}\bar{\rho}\tilde{M}_{1}r^{3} - M_{s}\right). 
\end{equation}
where $\delta_{D}$ represents the Dirac-delta function, and $r_{c}$
represents the lower limit for the subhalo survival: if a subhalo is 
located at a distance smaller than $r_{c}$,  they get completely 
destroyed by the strong tidal stripping effect.  The value of $r_{c}$ 
has been empirically found to be a few times the scale radius $r_{s}$ 
\citep{ghi-etal00,hay-etal03}. The upper limit $R_{2}$ in the
integration of $r$ is set from the expectation that the subhalos 
should be inside the virial radius of the host halo. 
Note that equations (\ref{eqn:cond}) and (\ref{eqn:localmf}) are 
all expressed in terms of the rescaled subhalo mass 
$\tilde{M}_{1}=M_{1}/M_{2}$. Their dependence on the host
halo mass comes only implicitly from $\gamma$ and $\xi$. 
It implies that the subhalo mass function should not depend strongly on 
the host halo mass, consistent with numerical detections 
\citep[e.g.,][]{delucia-etal03}. Furthermore, we compute the
contribution of the subhalos to the host halo mass by integrating 
equation (\ref{eqn:localmf}) over $M_{s}$, and find that it is 
approximately $10\%$ of the host halo mass, which is also consistent 
with numerical simulations \citep[e.g.,][]{tor-etal98}.
Finally, the global subhalo mass function is obtained by multiplying 
the local subhalo mass function by the host halo mass function and 
integrating it over the host halo mass: 
\begin{equation}
\label{eqn:globalmf}
\frac{dN(M_{s})}{d\ln M_{s}}d\ln M_{s}
= \int_{-\infty}^{\infty}d\ln M_{2}~
\frac{dN(M_{2})}{d\ln M_{2}}~
\frac{dN_{M_{2}}(M_{s})}{d\ln M_{s}}d\ln M_{s}, 
\end{equation}
where the host halo mass function $dN(M_{2})/d\ln M_{2}$ is given 
in equation (\ref{eqn:psmf}). 

\begin{figure}
\begin{center}
\plotone{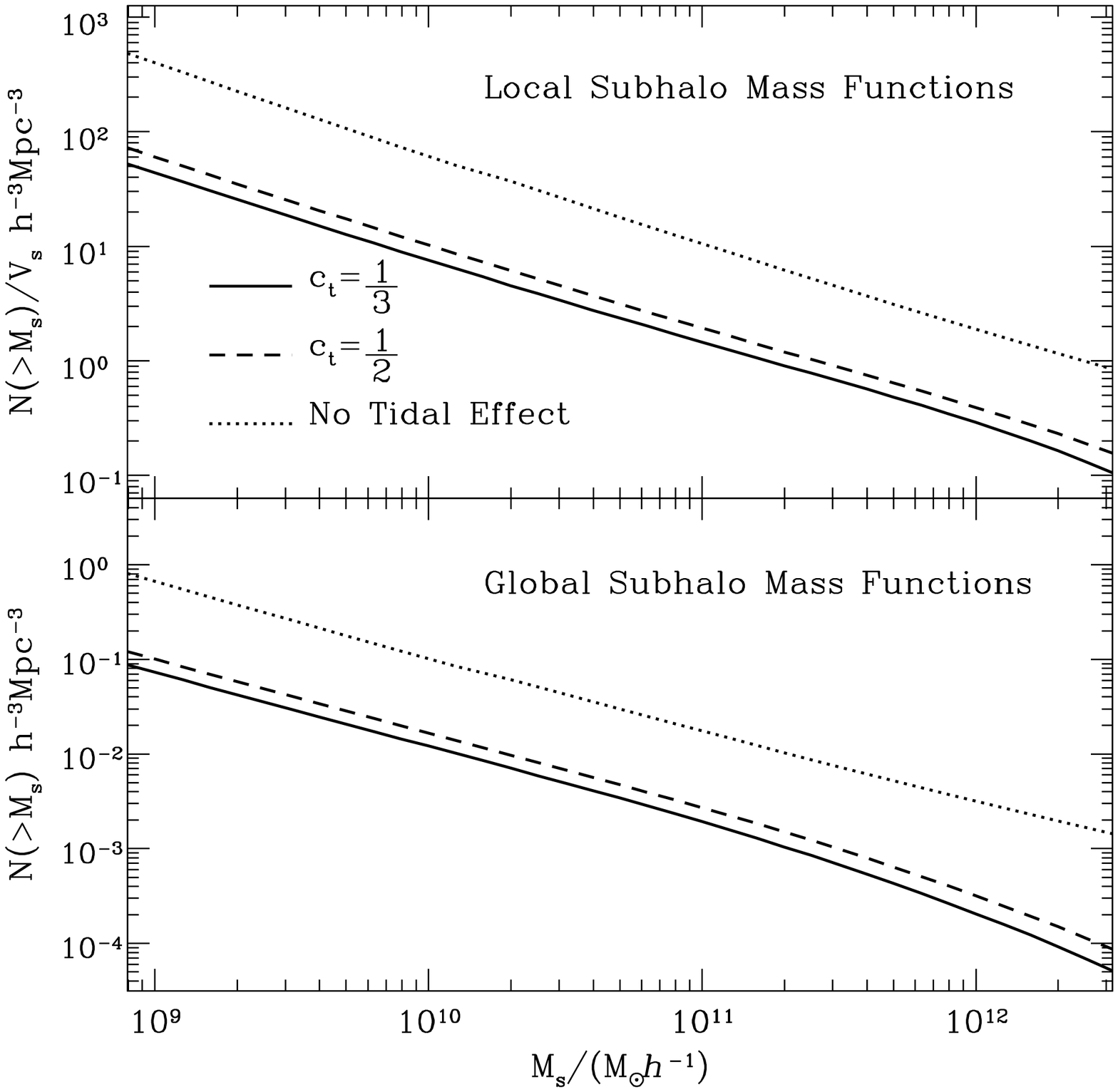} 
\caption{The mass functions of the dark halo substructures at redshift z=0:
{\it Upper Panel}: the local distribution and {\it Lower Panel}: the
 global distribution. 
The solid and dashed lines correspond to the two different values of
the free parameter $c_{t}$ in the the tidal survival condition 
(see, eq. [\ref{eqn:tsc}]).  The dotted lines correspond to the case of
 no tidal stripping effect.  
\label{fig:massft}}
\end{center}
\end{figure}

Figure \ref{fig:massft} plots the cumulative local (upper panel) and 
the global (lower panel) mass functions at redshift $z=0$ for the 
case of for a flat CDM cosmology \citep{bon-efs84} with the following 
choice of the cosmological parameters: the matter density $\Omega_{m}=0.3$, 
the vacuum energy density $\Omega_{\Lambda}=0.7$, the shape parameter 
$\Gamma=0.5$, the dimensionless Hubble constant $h=0.5$, and the rms 
density fluctuation on the scale of $8h^{-1}$Mpc, $\sigma_{8}=0.7$. 
For the local subhalo mass function, the host halo mass is chosen to be 
$M_{2} = 10^{14}M_{\odot}h^{-1}$. 
In each panel, the solid  and the dashed lines correspond to the cases 
of $c_{t} = 1/3$ and $c_{t} = 1/2$, respectively. For comparison, 
the subhalo mass function with no tidal effect is also plotted as dotted 
lines.  It is clear from Figure \ref{fig:massft} that both the global
and the local cumulative subhalo mass functions are close to a power law 
$N(M_{s}) \sim M^{l}_{s}$, and that the tidal mass-loss effect changes 
the subhalo mass function in a self-similar manner. 
We find that the power-law slope of the local subhalo mass function 
is $l \sim -0.8$ while that of the global subhalo mass function is 
slightly sharper $l \sim -0.9$. 
The power-law slope of the global subhalo mass function gets sharper 
in the high mass section. It is due to that massive subhalos 
experience the stronger tidal stripping effect (eq. [\ref{eqn:tsc}]) to 
lose more mass, and also that the dark halos which can afford to hosting 
such massive subhalos are rare whose number density decrease sharply 
(eq. [\ref{eqn:psmf}]).

\section{DISCUSSIONS AND CONCLUSIONS}

We provided for the first time a theoretical formalism in which one can 
estimate analytically the global and the local mass distribution of dark 
matter subhalos that undergo tidal mass-loss process. 
Adopting the simple tidal-limit approximation, 
we showed that the resulting mass functions are consistent with what has
been found in recent high-resolution N-body simulations, providing 
theoretical clues to the unique properties of the subhalo mass 
distribution: the power-law shape, weak dependence on host halo mass, 
self-similar change, and roughly  $10\%$ contribution of subhalos to 
the total mass.

Yet, it is worth mentioning that our subhalo mass functions are subject to
several caveats. The most obvious one is that we have oversimplified the 
subhao mass-loss process, using the simple tidal limit approximation, and 
also ignored the effects of dynamical frictions and close encounters 
between subhalos. It has been shown by numerical simulations that the 
tidal limit approximation underestimates the subhalo mass-loss
considerably \citep{hay-etal03}. 
Although it was shown by simulations that the most dominant
force that leads to the mass loss of the subhalos is the global tides
\citep{oka-hab99},  the dynamical frictions and subhalo close-encounters
may change the subhalo orbits making the subhalos
more susceptible to the tidal forces \citep{tor-etal98}.
We have also assumed simply that the subhalos rotate on stable circular 
orbits. However, in reality, the the subhalo orbits are quite eccentric,
changing with time \citep{tor-etal98,hay-etal03}.  Definitely, it will 
be quite necessary to refine our formalism by making more realistic 
treatments of the subhalo evolution, especially its mass-loss process.  

Finally, we conclude that our formalism is expected to provide an 
important first step toward realistic modeling of the abundance
distribution of dark halo substructures.  

\acknowledgments

We are very grateful to the anonymous referee who helped us greatly 
improve the original manuscript. 
We also acknowledge gratefully the research grant of the JSPS (Japan Society
of Promotion of Science) fellowship. This research was supported in part 
by the Grant-in-Aid for Scientific Research of JSPS (12640231).

\clearpage
\end{document}